\documentclass[twocolumn,aps,superscriptaddress, showkeys, nofootinbib,floatfix,linenumbers]{revtex4}
\usepackage{epsfig,bm,feynmf}
\usepackage{graphics}
\usepackage{amsmath}
\usepackage{mathrsfs}
\usepackage{appendix}
\usepackage{bm}
\usepackage[normalem]{ulem}  % \sout{old text} for strikeout
\usepackage[dvips]{color} % For blue in-text comments and additions

\renewcommand{\sout}{\bgroup \color{red} \ULdepth=-.5ex \ULset}

\begin{document}
\title{Azimuthal angle dependence of the longitudinal spin polarization in relativistic heavy ion collisions}

% authors
\author{Yifeng Sun}
\email{sunyfphy@physics.tamu.edu}
\affiliation{Cyclotron Institute and Department of Physics and Astronomy, Texas A$\&$M University, College Station, Texas 77843, USA}%

\author{Che Ming Ko}
\email{ko@comp.tamu.edu}
\affiliation{Cyclotron Institute and Department of Physics and Astronomy, Texas A$\&$M University, College Station, Texas 77843, USA}%

% date
\date{\today}% It is always \today, today,

\begin{abstract}
The azimuthal angle dependence of quark spin polarization in the longitudinal beam direction of non-central relativistic heavy ion collisions is studied in the chiral kinetic approach. Contrary to the prediction from models based on the assumption of thermal equilibrium of spin degrees of freedom that the quark spin polarization always points along the direction of local vorticity field, we find the two can have opposite directions due to the effect from the transversal component of vorticity field, which can lead to a redistribution of axial charges in the produced matter.  Our finding is consistent with the azimuthal angle dependence of the longitudinal spin polarization of $\Lambda$ hyperons, which is mainly determined by that of the strange quark, recently measured in the experiments by the STAR Collaboration at the Relativistic Heavy Ion Collider (RHIC).  
\end{abstract}

\maketitle

\section{introduction}

In non-central relativistic heavy ion collisions, a strong vorticity field is generated in the produced matter as a result of the large orbital angular momentum that is brought into the system.  This vorticity field can lead to the polarization of particles of non-zero spin along the direction of the vorticity field due to their spin-orbit~\cite{PhysRevLett.96.039901,PhysRevC.77.044902} or spin-vorticity coupling~\cite{PhysRevC.77.024906,Becattini20082452,Becattini201332,PhysRevC.94.024904}. Measurements of the global spin polarization of $\Lambda$ hyperons by the STAR Collaboration~\cite{STAR:2017ckg,PhysRevC.98.014910} have confirmed the existence of the most vortical fluid ever known, with an average vorticity of more than $10^{21}$ s$^{-1}$~\cite{PhysRevC.87.034906,PhysRevC.93.064907,PhysRevC.94.044910}. Theoretical studies based on the assumption of local thermal equilibrium~\cite{Karpenko2017,PhysRevC.94.054907,PhysRevLett.117.192301,PhysRevC.96.054908,PhysRevC.97.041902,PhysRevC.97.064902,Wei:2018zfb} or using the non-equilibrium chiral kinetic approach~\cite{PhysRevC.96.024906} have reproduced the experimental results. 

Besides its large global value, the vorticity field in the matter produced in relativistic heavy ion collisions also shows local structures~\cite{PhysRevC.94.044910,PhysRevC.97.041902,PhysRevLett.120.012302,PhysRevC.98.024905}.  An example of such local behavior is the quadrupole pattern in the distribution of the longitudinal component of vorticity field, which is along the beam direction, in the transverse plane of a heavy ion collision. According to Refs.~\cite{PhysRevLett.120.012302,PhysRevC.98.024905} based on the hydrodynamic or transport approach, the local variation of vorticity field can be verified by measuring the azimuthal  angle distribution of the spin polarization of $\Lambda$ hyperons because the latter is expected to always point along the direction of the vorticity field if one assumes that the spin degrees of freedom are in thermal equilibrium at the kinetic freeze-out of heavy ion collisions.  Although preliminary experimental results from the STAR Collaboration at RHIC~\cite{Niida:2018hfw} on the longitudinal spin polarization of $\Lambda$ hyperons in the transverse plane of heavy ion collisions have indeed shown a quadrupole pattern, its magnitude and sign are not consistent with those obtained from the longitudinal component of vorticity field calculated at the kinetic freeze-out in hydrodynamic or transport models.  In particular, the measured longitudinal spin polarization of $\Lambda$ hyperons points at an opposite direction from that of the vorticity field calculated from these models. Besides possible effects of initial hydro conditions, decays of higher lying states, and non-equilibration of spin degrees of freedom at kinetic freeze-out as suggested in Ref.~\cite{Becattini:2018lge}, the failure of these theoretical studies may also be due to the neglect of effects from the vorticity field on  quark propagations and scatterings.  Taking into account the redistribution of axial charges after propagations and scatterings of quarks in the vorticity field can lead to non-vanishing local axial chemical potential, which would further affect the polarization of quarks through their scatterings~\cite{PhysRevC.96.024906} and thus the resulting $\Lambda$ hyperons. As a result, this will modify the relation between the spin polarization of quarks or $\Lambda$ hyperons and the vorticity field  in models that assume local thermal equilibrium but neglect the redistribution of axial charges.  In the present study, we use the chiral kinetic approach to take into account the different effects of vorticity field on quarks with positive and negative helicities, and show that it can indeed explain the different azimuthal angle distributions between the longitudinal spin polarization of $\Lambda$ hyperons and the longitudinal component of the vorticity field in the produced matter as observed in experiments.

This paper is organized as follows. In the next section, we show how a specific transverse vorticity field can lead to the redistribution of axial charges, and how this can give rise to a quadrupole pattern in the spin polarization of quarks along the longitudinal direction. A brief description of the chiral kinetic approach is given in Sec. III. In Sec. IV, with initial conditions taken from the event generator AMPT model~\cite{PhysRevC.72.064901}, we solve the chiral kinetic equations of motion for quarks and their scatterings in the presence of vorticity field, which is calculated locally, to study the above effect by including one component of the vorticity field.  We then show that under certain conditions the combined effects due to the contributions from all components of the local vorticity field can explain the experimental results. Finally, a summary is given in Sec. V.

\section{Effect of transverse vorticity field on  longitudinal spin polarization}\label{spin}

\begin{figure}[h]
\centering
\includegraphics[width=1\linewidth]{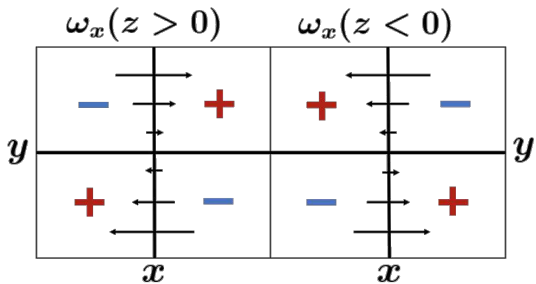}
\caption{(Color online) Distribution of the in-plane $x$ component of the vorticity field $\omega_x$~\cite{PhysRevC.98.024905} and the induced redistribution of axial charges in the transverse plane of a heavy ion collision.}
\label{vorticity}
\end{figure}

Since the matter produced in relativistic heavy ion collisions mostly locates around the center of the collisions, i.e., $z=0$ on the longitudinal coordinate, the velocity in the transverse direction, which is driven by the pressure in the matter, has a maximum value at $z=0$ and decreases with increasing $|z|$. This then results in a transverse vorticity field,  $\boldsymbol{\omega}_\perp=\frac{1}{2}\partial_zv_\perp(r,z)\mathbf{e}_\phi$, that is clockwise in the region of $z>0$ and anti-clockwise in the region of $z<0$ as shown by the arrows in Fig.~\ref{vorticity}~\cite{PhysRevC.98.024905}.  In the above, $v_\perp(r,z)$ and $\mathbf{e}_\phi$ are, respectively, the velocity field and unit vector in the transverse plane, with the latter making an angle $\phi$ with respect to the reaction plane.  For the in-plane $x$ component of the transverse vorticity field, it leads to not only the spin polarization of quarks along the $x$ direction but also an axial charge current in that direction, i.e., $\mathbf{j}^A\propto \frac{T^2}{6}\boldsymbol{\omega}$ with $T$ denoting the temperature of the quark matter, due to the so-called gravitational anomaly~\cite{PhysRevLett.107.021601}. 

In the region of $z>0$ shown in the left panel of Fig.~\ref{vorticity}, the axial charge current is thus along the positive $x$ direction for $y>0$ and along the negative $x$ direction for $y<0$, resulting in an quadrupole pattern in the distribution of axial charges.  Therefore, there are more right-handed than left-handed quarks in the quadrants $xy>0$ and more left-handed than right-handed quarks in the quadrants $xy<0$ as shown by the "$+$" and "$-$" symbols in the left panel of Fig.~\ref{vorticity}. Because the momenta of quarks in the region of $z>0$ are preferentially along the positive $z$ direction, their longitudinal spin polarizations are thus positive in the quadrant $xy>0$ and negative in the quadrant $xy<0$.  For the longitudinal spin polarizations of quarks in the region of $z<0$, they show a similar pattern as those in the region of $z>0$ since the momenta and axial charges of quarks in the region of $z<0$ are opposite to those in the region $z>0$. This azimuthal angle dependence of quark longitudinal spin polarizations agrees with the experimental results from the STAR Collaboration for those of $\Lambda$ hyperons. As for the effect of global vorticity on the quark global spin polarization~\cite{PhysRevC.96.024906}, the effect of the transverse component of local vorticity on the local quark  spin polarization will be further enhanced if one includes the effect of spin-vorticity coupling in the scattering of quarks.

The above effect due to the $x$ component of the vorticity field is, however, reduced by the $y$ component of the vorticity field~\cite{PhysRevC.98.024905}. This is because the latter is along the negative $y$ direction for $x>0$ and along the positive $y$ direction for $x<0$ in the region of $z>0$, which thus leads to more right-handed than left-handed quarks in the quadrants $xy<0$ for the region $z>0$.  If both the produced quark matter and the vorticity field have rotational symmetry around the $z$ axis, there will be an exact cancellation between the effects from the $x$ and $y$ components of the vorticity field. Since this is not the case in non-central heavy ion collisions, the cancellation is not perfect, and there remains an additional contribution to the longitudinal spin polarization of quarks from the transverse components of the vorticity field besides its longitudinal component. To study this effect quantitatively, we use in the present study the chiral kinetic approach to describe the dynamics of quarks under the influence of a self-consistently calculated vorticity field with the initial quark distributions from the AMPT model~\cite{PhysRevC.72.064901}, which is an event generator for heavy ion collisions at ultra-relativistic energies. 

\section{The chiral kinetic approach}

In the chiral kinetic approach to massless quarks and antiquarks in an vorticity $\boldsymbol\omega$ field, their equations of motion are given by~\cite{Sun:2016mvh,PhysRevLett.109.162001,PhysRevD.96.016002,PhysRevC.96.024906}
\begin{eqnarray}\label{chiral}
&&\dot{\mathbf{r}}=\frac{\hat{\mathbf{p}}+2\lambda p(\hat{\mathbf{p}}\cdot\mathbf{b})\boldsymbol{\omega}}{1+6\lambda p(\mathbf{b}\cdot\boldsymbol{\omega})},\quad\dot{\mathbf{p}}=0,
\end{eqnarray}
where $\lambda=\pm 1$ is the helicity of quark or antiquark, and $\mathbf{b}=\frac{\mathbf{p}}{2p^3}$ is the Berry curvature that results from the adiabatic approximation of taking the spin of a massless parton to be always parallel or anti-parallel to its momentum. Corrections to above equations due to finite quark masses ($m_u=3$ MeV, $m_d=$ 6 MeV, and $m_s=100$ MeV) can be included by replacing $\hat{\mathbf{p}}$, $p$ and $\mathbf{b}$ with $\frac{\mathbf{p}}{E_p}$, $E_p$ and $\frac{\hat{\bf{p}}}{2E_p^2}$, respectively, as in Ref.~\cite{PhysRevD.89.094003}. 

The factor $\sqrt{G}=1+6\lambda p(\mathbf{b}\cdot\boldsymbol{\omega})$ in the denominator of Eq.(\ref{chiral}) modifies the phase-space distribution of partons and ensures the conservation of vector charge.  The modified parton equilibrium distribution can be achieved from parton scatterings by requiring the parton momenta $\mathbf{p}_3$ and $\mathbf{p}_4$ after a two-body scattering, which is determined by their total scattering cross section, with the probability $\sqrt{G(\mathbf{p}_3)G(\mathbf{p}_4)}$~\cite{PhysRevC.96.024906}.  

As shown in Ref.~\cite{PhysRevC.96.024906} and also mentioned previously, the vorticity field through the phase-space factor $\sqrt{G}$ has a larger effect on the spin polarization of quarks from their scatterings than propagations.  This is because including $\sqrt{G}$ in scattering makes quarks of positive $p_y$, i.e., momentum in the $y$ direction perpendicular to the reaction plane of a heavy ion collision, more right-handed and quarks of negative $p_y$ more left-handed, resulting quickly in a net spin polarization.  For the effect of chiral equations of motion on quark propagations, although it can  lead to an axial current in coordinate space, resulting in more right-handed quarks in the upper hemisphere and left-handed quarks in the lower hemisphere of the reaction plane, it takes additional times for quarks to have positive $p_y$ in the upper hemisphere and negative $p_y$ in the lower hemisphere, so that quarks can acquire finite spin polarization.  

\section{Longitudinal spin polarization}

For the application of the chiral kinetic equations of motion to heavy ion collisions, we use the phase-space distributions of quarks and antiquarks from the AMPT model without the partonic and hadronic evolutions as the initial conditions. Assuming that there is no local topological charges in the produced quark matter, the helicity of each quark  is determined randomly with equal probabilities to be positive and negative.  The local vorticity field is calculated from the flow field $\mathbf{u}=\gamma \mathbf{v}=\frac{1}{\sqrt{1-v^2}}\mathbf{v}$ according to $\boldsymbol{\omega}=\frac{1}{2} \boldsymbol{ \nabla} \times \mathbf{u}$.  For the flow field, it is obtained from averaging the velocities of quarks in a local cell of size of dimensions $\Delta x=\Delta y=0.5$ fm and $\Delta \eta=0.2$, where $\eta=\frac{1}{2}\ln\frac{t+z}{t-z}$ is the space-time rapidity.  Since quarks from the AMPT model are produced at different formation times, those that are not formed at a given time are propagated back to that time and included in the evaluation of the flow field.  As to the quark scattering cross section, we use a constant value of 10 mb and an isotropic angular distribution for simplicity.   Results given below are obtained from the phase-space distributions of quarks after they stop scattering in Au+Au collisions at $\sqrt{s_{NN}}=200$ GeV with their initial distributions taken from the AMPT model at the impact parameter $b=8.87$ fm to simulate the 30-40\% centrality in the STAR experiments. 

\subsection{Time evolution of local vorticity field}\label{local}

\begin{figure}[h]
\centering
\includegraphics[width=1\linewidth] {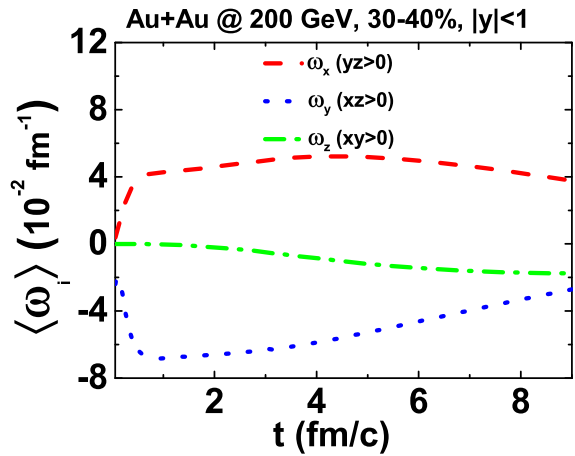}
\caption{(Color online) Time evolution of different components of the average vorticity of midrapidity quarks in  selected regions in the transverse plane.}
\label{wxyz}
\end{figure}

To see the local structure of the vorticity field, we evaluate the average value of its components $\omega_x$, $\omega_y$, and $\omega_z$ in different spatial regions according to 
\begin{eqnarray}
&&\langle \omega_i\rangle=\frac{\sum_{j} \omega_i(x_j,y_j,z_j)}{\sum_j 1}
\end{eqnarray}
with $i=x, y, z$ and $j$ running over all midrapidity quarks in selected spatial regions such as $yz>0$, $xz>0$ and $xy>0$. Shown in Fig.~\ref{wxyz} are the time evolutions of $\langle\omega_x\rangle$ along the  $x$ direction in the region $yz>0$, and $\langle\omega_y\rangle$ and $\langle\omega_z\rangle$ along the $y$ and $z$ directions in the regions $xz>0$ and $xy>0$, respectively.  It is seen that $\langle\omega_x\rangle$ is positive and $\langle\omega_y\rangle$ and $\langle\omega_z\rangle$ are negative, with their final values similar to those from Ref.~\cite{PhysRevLett.120.012302} based on the hydrodynamic model at kinetic freeze-out and Ref.~\cite{PhysRevC.98.024905} using the AMPT model after including both partonic and hadronic evolutions. Figure~\ref{wxyz} also shows that the magnitudes of $\omega_x$ and $\omega_y$ increase faster with time and are much larger than the magnitude of $\omega_z$.  We note that the initial non-zero value of $\omega_y$ is due to the large global orbital angular momentum in the $y$ direction in non-central heavy ion collisions.

\subsection{Time evolution of the longitudinal spin polarization of quarks}

\begin{figure}[h]
\centering
\includegraphics[width=1\linewidth] {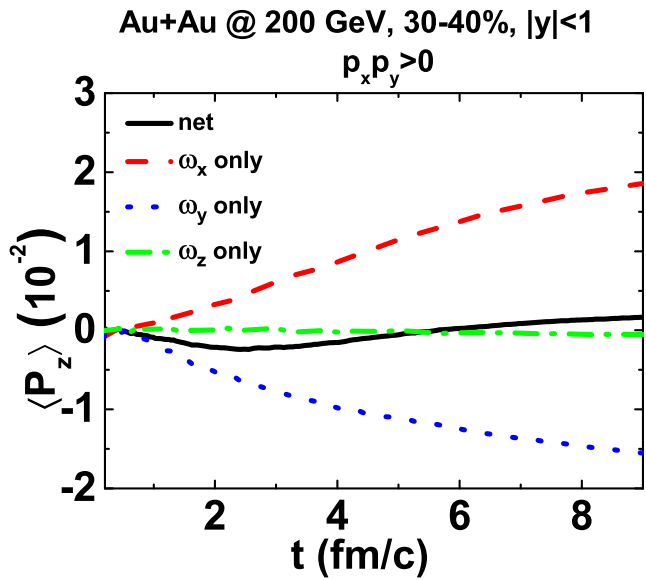}
\caption{(Color online) Time evolution of average longitudinal spin polarization of midrapidity quarks with momenta satisfying $p_xp_y>0$.}
\label{pz}
\end{figure}

Because of the local structure of vorticity field shown in Fig.~\ref{wxyz}, quarks of momenta $p_xp_y>0$ acquires a 
longitudinal spin polarization in the $z$ direction. Including only the $\omega_x$, we find that the average longitudinal spin polarization of these quarks is along the positive $z$ direction and increases with time, which is shown by the red dashed line in Fig.~\ref{pz} and confirms the expectation discussed in Sec.~\ref{spin}. The resulting magnitude of the spin polarization due to $\omega_x$ is of the order of 10$^{-2}$ and is appreciable. The longitudinal spin polarization of quarks of momentum $p_xp_y>0$ due to $\omega_y$ also increases with time but is along the negative $z$ direction, again agreeing with the expectation discussed in Sec.~\ref{spin}. Its final magnitude is also of the order of 10$^{-2}$.

Since $\omega_z$ is along the negative $z$ direction in the region $xy>0$, it leads to a longitudinal spin polarization in the negative $z$ direction for quarks of momenta $p_xp_y>0$, as shown by the green dash-dotted line in Fig.~\ref{pz}. However, its magnitude is only of the order of 10$^{-3}$ and slowly increases with time. 

Including all components of the vorticity field, which is shown by the black solid line in Fig.~\ref{pz}, we find that the total longitudinal spin polarization of quarks of momenta $p_xp_y>0$ is initially along the negative $z$ direction, as a result of the larger effect of $\omega_y$ than that of $\omega_x$.  After about 2.5 fm$/c$, the effect of $\omega_x$ becomes more important than that of $\omega_y$, and this makes the longitudinal spin polarization of these quarks less negative. Finally, the sign of the longitudinal polarization is along the positive $z$ direction after 5 fm$/c$ when the effect of $\omega_x$ dominates over the combined effects of $\omega_y$ and $\omega_z$.

\subsection{Rapidity dependence of  longitudinal spin polarization}

\begin{figure}[h]
\centering
\includegraphics[width=1\linewidth] {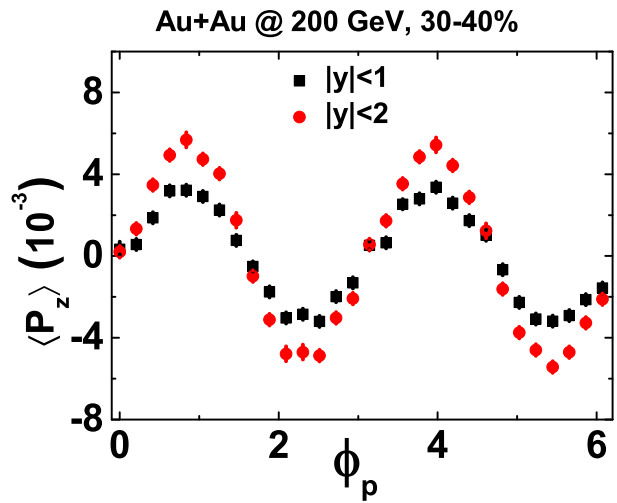}
\caption{(Color online) Average longitudinal spin polarization of quarks as a function of azimuthal angle $\phi_p$ for  different rapidity ranges.}
\label{phipz}
\end{figure}

In Fig.~\ref{phipz}, we show the longitudinal spin polarization of quarks as a function of the azimuthal angle in the transverse plane of heavy ion collisions for different rapidity ranges. It is seen that the longitudinal spin polarization indeed has a quadrupole pattern and is positive for quarks $p_xp_y>0$, which has the same pattern and similar magnitude as those of $\Lambda$ hyperons measured in experiments~\cite{Niida:2018hfw}, and differs from the longitudinal polarization calculated from $\omega_z$ by assuming local thermal equilibrium of the spin degrees of freedom. Furthermore, the amplitude of the azimuthal dependence, which can be expressed as $\rm{sin}(2\phi_p)$, is larger for the larger rapidity, and this is due to the larger values of longitudinal and transverse vorticities at larger $\eta$~\cite{PhysRevC.94.044910,PhysRevC.96.054908}.  

\begin{figure}[h]
\centering
\includegraphics[width=1\linewidth] {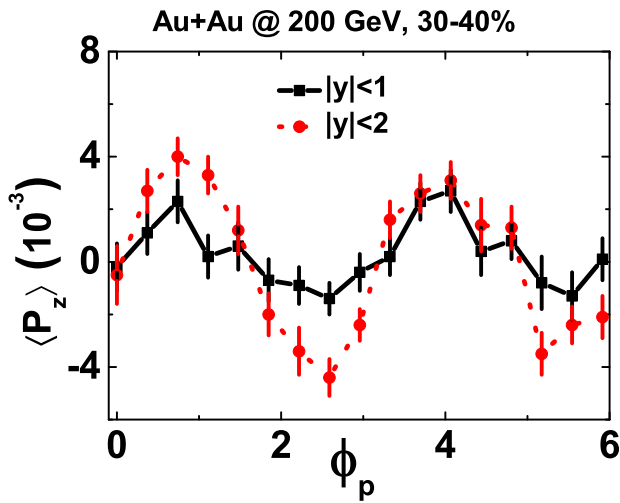}
\caption{(Color online) Same as Fig.~\ref{phipz} for strange quarks only.}
\label{phipzs}
\end{figure}

We also show the longitudinal spin polarization of strange quarks in Fig.~\ref{phipzs}, which is expected to be almost identical to that of $\Lambda$ hyperons~\cite{PhysRevLett.96.039901,PhysRevC.96.024906,PhysRevC.97.034917}. It is seen that the amplitude of the azimuthal angle dependence of the longitudinal spin polarization of strange quarks is smaller than that of light quarks, but is still  comparable to the experimental results~\cite{Niida:2018hfw}. The reason for this is because of the mass effect in the chiral kinetic approach and the different spatial and temporal distributions between initial strange and light quarks from the AMPT model.  

We further find that with a smaller quark cross section, the longitudinal spin polarization of quarks would decrease and can even change the overall sign of the quadrupole pattern of the longitudinal spin polarization. This thus indicates that taking into account the non-equilibrium effect, which is included in the chiral kinetic approach, is important for understanding the local spin polarization of quarks and thus $\Lambda$ hyperons.

\section{Summary}

Using the chiral kinetic approach, which takes into account the axial charge redistribution in the vorticity field, with initial quark phase-space distributions taken from the AMPT model, we have studied the effect of the transverse components of local vorticity field on the longitudinal spin polarization of quarks. We have found that the longitudinal spin polarization of quarks depends not only on the longitudinal component of the vorticity field but also on its transverse components.  Using a constant quark scattering cross section of 10 mb, we have obtained a longitudinal spin polarization of quarks that has an azimuthal angle dependence and amplitude similar to those measured in experiments for $\Lambda$ hyperons, as a result of the dominant effect of the in-plane component $\omega_x$ over those of the out-of-plane component $\omega_y$ and the longitudinal component $\omega_z$ of local vorticity field. We have also found that decreasing the quark scattering cross section leads to a reduction of the longitudinal spin polarization in the positive $z$ axis along the beam direction. Our study thus demonstrates the importance of non-equilibrium effects and the local structure and time evolution of vorticity field on the spin polarizations of quarks in relativistic heavy ion collisions. However, although the $\Lambda$ longitudinal spin polarization has a similar azimuthal angle dependence as that of strange quarks immediately after hadronization of the quark matter, how it is affected during the hadronic evolution  needs to be studied to see if its observed opposite direction to that of local vorticity field is indeed due to the mechanism described in the present study.

\section*{ACKNOWLEDGEMENTS}

This work was supported in part by the US Department of Energy under Contract No. DE-SC0015266 and the Welch Foundation under Grant No. A-1358.

\bibliography{ref}

\end{document}